\RequirePackage{fixltx2e}
\documentclass[aip,jcp,reprint,showpacs]{revtex4-1}

\usepackage{amsmath,amsfonts}
\usepackage{graphicx}
\usepackage[acronym]{glossaries}
\usepackage{dcolumn}
\usepackage{bm}
\usepackage{ifpdf}
\usepackage{natbib}
\usepackage{wasysym}
\usepackage[byname]{smartref}
\usepackage{color}
\usepackage{lipsum}
\usepackage{nicefrac}
\usepackage[breaklinks, colorlinks, linkcolor=blue, citecolor=blue, urlcolor=blue]{hyperref}
\usepackage[table]{xcolor}
\usepackage{multirow}
\usepackage{braket}
\usepackage{float}
\usepackage{comment}
\usepackage{enumerate}
\usepackage{enumerate}

\usepackage{amsmath,amssymb,graphicx}
\usepackage{epsfig}

\newcommand{\abs}[1]{\left| #1 \right|} 
\renewcommand{\d}[2]{\frac{d #1}{d #2}} 

\DeclareMathOperator{\Tr}{Tr}
\newcommand{\beq}{\begin{equation}}
\newcommand{\eeq}{\end{equation}}

\newcommand{\bea}{\begin{eqnarray}}
\newcommand{\eea}{\end{eqnarray}}

\begin{document}


\title{Can Seebeck coefficient identify quantum interference in molecular conduction?}

\author{Lena Simine}
\affiliation{Chemical Physics Theory Group, Department of Chemistry, University of Toronto,
80 Saint George St. Toronto, Ontario, Canada M5S 3H6}

\author{Wei Jia Chen}
\affiliation{Chemical Physics Theory Group, Department of Chemistry, University of Toronto,
80 Saint George St. Toronto, Ontario, Canada M5S 3H6}

\author{Dvira Segal}
\email{dsegal@chem.utoronto.ca}
\affiliation{Chemical Physics Theory Group, Department of Chemistry, University of Toronto,
80 Saint George St. Toronto, Ontario, Canada M5S 3H6}

\date{\today}
\begin{abstract}
We look for manifestations of quantum interference effects in the Seebeck coefficient of
a molecular junction, when the electronic conductance exhibits pronounced destructive interference features
due to the presence of quasi-degenerate electronic states which differ in their spatial symmetry.
We perform our analysis by considering three separate limits
for electron transport: coherent, fully dephased, and suffering inelastic scattering with molecular vibrations.
We find that while the conductance displays strong signatures of the underling transport mechanisms:
destructive quantum interference features in the coherent case and thermal activation characteristics in the inelastic limit,
the Seebeck coefficient conceals details of electron dynamics while it robustly reveals information about
the energy characteristics of the junction.
We provide closed-form expressions for the
electronic conductance and the thermopower of our system as a function of temperature, gate voltage and hybridization
energy in different transport limits,
then exemplify our analysis on a specific conjugated molecule with quasi-degenerate orbitals of different spatial symmetry.
\end{abstract}


\maketitle

\section{Introduction}

Quantum interference (QI) effects in single molecule junctions were recently explored
experimentally \cite{LathaQI,MolenQI,Zant,thossE}
and theoretically \cite{gemmaJCP08, gemmajacs,gemmaBeil,thossT,novel,MarkussenN, MarkussenPCCP,
Markussen}, identifying different classes of molecules
which may exhibit this behavior even at a room temperature in ambient conditions.
For example, a benzene ring coupled to the rest of the molecule in a meta-configuration
supports low conductance values relative to the para-coupled configuration \cite{Zant}.
This observation can be explained from the behavior of the transmission function for $\pi$ electrons in the system:
For meta-configurations it has a node close to the Fermi energy, the result of
destructive interference between  HOMO and LUMO molecular orbitals. This `anti-resonance' feature is missing
in the para situation.
Similarly, acyclic cross-conjugated molecules demonstrate anti-resonance
QI features near the Fermi energy \cite{MolenQI,gemmajacs,gemmaBeil}.

A different class of molecules with quasi-degenerate electronic levels close
to the Fermi energy was examined experimentally in
Ref. \cite{thossE} with the 2,2'- dimethylbiphenyl (DMBP) serving as a case study for this family of molecules.
In this system, the two benzene rings are almost orthogonal in orientation,
providing a weak coupling energy $\Delta$ between the $\pi$ systems on the two rings.
In the picture of molecular orbitals, weak electronic coupling corresponds to a small
energy splitting $\Delta$ between the states of opposite $L-R$ symmetry.
These orbitals are termed ``quasi-degenerate" if their broadening due to the
coupling to the metals is larger than level splitting $\Delta$.
In this situation, electrons cross the molecule
through two identical ``paths" with a $\pi$ phase difference, resulting in (destructive) QI effects and the suppression of
the transmission function, relative to the case with no interference.
QI was deduced experimentally in such molecules from the enhancement of the current with increasing temperature
or under stretching, interpreted as the suppression of destructive interference due
to vibration-induced electronic decoherence in the former case, or
the breakdown of levels' degeneracy, the result of stretching.

Mechanisms of electron transport in single molecules,
coherent-elastic motion and thermally activated dynamics \cite{ReviewN},
were diagnosed from the temperature dependence of the current-voltage characteristics \cite{Selzer},
inelastic electron tunneling spectroscopy signals \cite{IETS} and
Raman response of molecules under bias \cite{Raman}.
Complementing conductance measurements, the thermopower, also referred to as the Seebeck
coefficient, was suggested as an  independent probe to determine mechanisms
of molecular electronic conduction.

The Seebeck coefficient is a linear response quantity.
It measures the voltage $\Delta V$ which develops when a small temperature difference is applied
under the condition that the net charge current vanishes, $S\equiv -\frac{\Delta V}{\Delta T}|_{I_e=0}$.
In the coherent tunneling limit, for largely separated HOMO-LUMO orbitals, 
the sign of the current (thus the Seebeck coefficient) discloses the main charge carries in the system:
The Seebeck coefficient is negative for LUMO (electron) conducting junctions, and
positive for HOMO (holes) dominated conduction \cite{datta,Wang}.
Theoretical studies have further analyzed the effect of molecular length, conformation, and chemical content
on the thermopower \cite{cuevas, Chen,ChenJPCC,Louie,Barry, gemmaL,baranger,cuevasdiss},
by combining the Landauer formalism, a single-particle method, with
ab initio electronic structure calculations.
Other studies emphasized the
sensitivity of $S$ to fluctuations in molecular structure \cite{dubi1,dubi2}.

Inelastic processes may affect the Seebeck coefficient in molecular junctions and other nanoelectromechanical
systems, altering its length and temperature dependence, compared to the coherent case
\cite{Koch,Segal,Kamil,Galp,Ora, ChenVib,ChenVibn,polaron,Lili,Cata,deph,Baowen}.
Means to enhance the thermopower using quantum coherent effects were explored in Refs.
\cite{Staff9,Staff10, Wierzbicki, Asai,wacker, Trocha, cuniberti},
by making use of quantum resonance and interference effects.
Theoretical and experimental studies of thermoelectricity in molecular junctions were reviewed in  Ref. \cite{dubiRev}.
More broadly,  approaches for calculating the thermopower and heat to work conversion efficiency
in the context of non-equilibrium thermodynamics were recently described in Ref. \cite{casati}.

STM break junction measurements of the Seebeck coefficient in single molecules were reported
in e.g., Refs. \cite{ReddyS,MalenE08,Reddy9,Malen9,Reddy10,Reddy11,fullerene,Tao,Agrait, Reddy14}, analyzing its behavior
as a function of molecular length, constituents, intermolecular interactions and gate electrode. These studies
identified orbital hybridization, contact-molecule energy coupling and geometry, and whether
the conductance is HOMO or LUMO dominated.
Particularly, recent experiments allowed a simultaneous measurement of the electronic conductance and the thermopower,
providing critical tests for transport theories:
while the conductance decreases exponentially with molecular length in the tunneling regime, the thermopower
increases (possibly nonlinearly) with size \cite{Latha12,Latha13}, in agreement with theoretical predictions
\cite{Segal,cuevas}.

In this work, we look for signatures of quantum interference effects in the thermopower,
in situations when such effects strongly affect the electronic conductance.
We focus on a class of molecules
with quasi-degenerate orbitals of different $L$-$R$ symmetry,
for example, the DMBP molecule, and consider resonance and off-resonance
(gate controlled) situations under coherent, fully dephased, and inelastic transport mechanisms.
We ask the following question:
Can the thermopower pinpoint on transport mechanisms and QI effects? 
Interestingly, the answer is negative for the geometry considered in this work, in a broad range of parameters. Particularly, at room temperature
we find that the Seebeck coefficient conceals underlying electron dynamics in the junction,
providing identical features for coherent and inelastic transport situations.
This observation connects with the thermodynamic interpretation of
the thermopower; it reflects the entropy per particle, but misses dynamical information \cite{Shastry}.

\begin{figure}[htbp]
\vspace{0mm} {\hbox{\epsfxsize=80mm \hspace{0mm}\epsffile{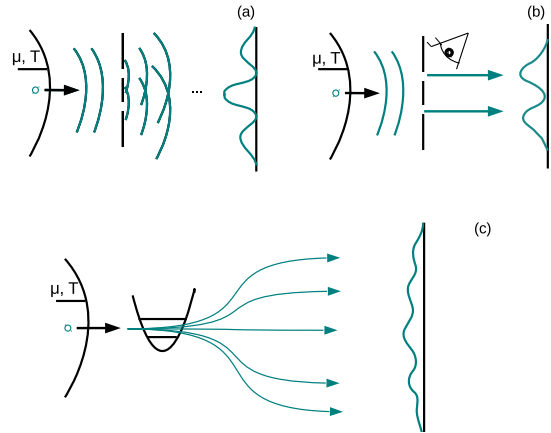}}}
\caption{\normalsize{Scheme of  different transport limits explored in this work.
(a) Coherent transmission of an electron (represented by a circle at the left metal)
through two closely-spaced electronic levels, represented by a two-slit interferometer.
(b) Transmission of an electron through two closely-spaced electronic levels while suffering dephasing
due to the presence of a dephasing agent, an observer (eye).
(c) Inelastic transmission: incoming electrons exchange energy with a molecular vibration.}}
\label{schemeP}
\end{figure}

We explore manifestations of QI in the conductance and the thermopower
using a minimal two-orbital model to describe the molecular object.
Assuming that these molecular orbitals are closely spaced,
we consider the following three situations, schematically depicted in Fig. \ref{schemeP}.
In panels (a)-(b) we sketch the molecule in the language
of Young's double-slit experiment: electrons are ``shot" from
the left lead, pass through the slits, and are then collected in the right lead.
The coherent case is shown in panel (a) with quantum interference effects dispersing the particles.
Panel (b) includes an observer which induces dephasing, suppressing interference patterns
to reach the familiar classical double-maxima curve.
In the third scenario,  presented in panel (c), electrons exchange energy with molecular vibrations and
they are inelastically scattered into states with energies
determined by the frequency of the vibrational mode.
%
We contrast transport results from these three limits, to explore signatures of QI in the thermopower.

The paper is organized as follows. In Sec. \ref{Smodel} we present our model and its
three limits (coherent, dephased, inelastic).
In Sec. \ref{Smethod} we explain the calculation of the  conductance and thermopower under each mechanism.
Sec. \ref{Sresult} includes numerical results, Sec. \ref{Ssummary} concludes.
%

\begin{figure}[htbp]
\vspace{0mm} {\hbox{\epsfxsize=80mm \hspace{0mm}\epsffile{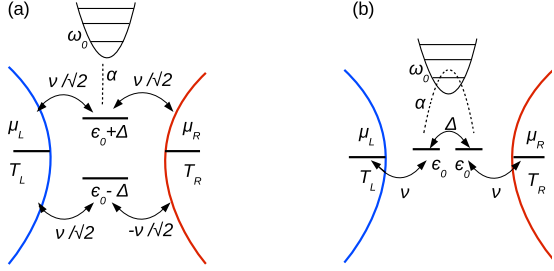}}} 
\caption{\normalsize{Scheme of the molecular junction under consideration.
(a) Molecular orbital representation: Electrons in the higher orbital interact with a vibrational mode.
(b) After transformation to the local site basis we reach the ``donor-acceptor" model. In this picture
electrons can tunnel between the two sites (strength $\Delta$) and the vibrational mode interacts
with electrons in the molecules with diagonal (population) and off-diagonal (tunneling processes)
terms.
Double arrows depict electron tunneling processes, dashed lines describe electron-vibration interactions.
}}
\label{schemeM}
\end{figure}

\section{Model}
\label{Smodel}

The Hamiltonian for a molecule between two leads is written as
\bea
H_{MO} = H_M + H_{el} + V_{M-el}.
\label{eq:Htot}
\eea
The molecule $H_M$ is described by molecular orbitals (MO) and a particular vibrational
degree of freedom which is linearly
coupled to electron densities in the molecule. We include only two such orbitals,
e.g., the HOMO and LUMO in Fig. \ref{schemeM}(a),
denoted by '1' and '2', respectively,
\bea
H_M &=&
(\epsilon_0-\Delta)c_1^{\dagger}c_1
+ (\epsilon_0+\Delta)c_2^{\dagger}c_2
\nonumber\\
&+& \hbar\omega_0b_0^{\dagger}b_0 +
\alpha c_2^{\dagger}c_2\left(b_0^{\dagger} + b_0\right).
\label{eq:HM}
\eea
The operator $c_n^{\dagger}$ ($c_n$) creates (annihilates) an electron on the $n=1,2$ molecular state,
placed symmetrically around the energy $\epsilon_0$ with an energy shift $\pm \Delta$.
Coulomb interaction terms are disregarded. 
The specific vibrational mode of frequency $\omega_0$ (creation operator $b_0^{\dagger}$) is assumed harmonic,
and for simplicity, it is only coupled to charge density on orbital 2 with the coupling strength $\alpha$.
Given that the total Hamiltonian does not commute with this interaction term,
this coupling allows for both electron dephasing and energy exchange processes between electrons
and the vibrational degree of freedom.
The two metal leads $\nu=L,R$ consist noninteracting electrons
\bea
H_{el}  &=& \sum_{k\in L,R} \epsilon_k c_k^{\dagger}c_k.
\label{eq:Hel}
\eea
Here $c_k^{\dagger}$ ($c_k$) creates (annihilates) an electron of energy $\epsilon_k$ in the electrode.
The molecule-metal electronic tunneling terms are given by
\bea
V_{M-el} \,\,&=&\,\, \sum_{r\in R}\frac{\upsilon_r}{\sqrt{2}}(c_r^{\dagger}c_2 - c_r^{\dagger}c_1 + h.c.)
\nonumber\\
&+& \sum_{l\in L}\frac{\upsilon_l}{\sqrt{2}}(c_l^{\dagger}c_2 + c_l^{\dagger}c_1 + h.c.).
\label{eq:VMel}
\eea
The couplings $\upsilon_k$ are assumed to be real valued.
The model is referred to as the ``DES (destructive) model" in the literature \cite{thoss1}, see Fig. \ref{schemeM}(a).
The negative sign, $e^{i\pi}$, corresponds to a phase difference between paths,
resulting in a perfect destructive interference of the transmission function in the absence of vibrations,
when the molecular electronic levels are degenerate, $\Delta=0$, as we discuss below in more details.
This phase difference corresponds to orbitals of different symmetry: Orbital 2  has an even (gerade) symmetry
with respect to the $L- R$  symmetry axis, orbital '1' has an odd (ungerade) symmetry
with respect to this axis.

In the atomic orbital (AO) representation, $c_d=\frac{1}{\sqrt 2}(c_1+c_2)$ and $c_a=\frac{1}{\sqrt 2}(c_2-c_1)$,
the Hamiltonian $H_{MO}$  
translates to a two-site donor-acceptor junction, see Fig. \ref{schemeM}(b),
\bea
H_{AO}&=&
\epsilon_0n_a + \epsilon_0n_d +  \Delta (c_a^{\dagger}c_d + c_d^{\dagger}c_a)
\nonumber\\
&+&  \sum_{k \in L,R} \epsilon_kc_k^{\dagger}c_k  + \hbar\omega_0 b_0^{\dagger}b_0
\nonumber\\
&+&
\sum_{l\in L}\upsilon_l(c_l^{\dagger}c_d + c_d^{\dagger}c_l)+
\sum_{r\in R}\upsilon_r(c_r^{\dagger}c_a + c_a^{\dagger}c_r).
\nonumber\\
&+&
\frac{\alpha}{2} \left[ \lambda_{di}(n_d + n_a)
+ \lambda_{o}
  (c_a^{\dagger}c_d + c_d^{\dagger}c_a) \right]
(b_0^{\dagger} + b_0).
\nonumber\\
\label{eq:Hda}
\eea
In this picture, $\Delta$ stands for the tunneling energy between donor and acceptor sites and
the perfect destructive interference pattern at $\Delta=0$
corresponds to a disconnected junction which naturally cannot transfer electrons between the metals.
Here $n_d=c_d^{\dagger}c_d$ and $n_a=c_a^{\dagger}c_a$ are number operators for the $d$ and $a$ orbitals.
The electron-vibration coupling energy  $\alpha$ is dressed
here by the flags $\lambda_{di}$ and $\lambda_{o}$, taking the values $0,1$, allowing us to discern different
effects:
$\lambda_{di}$ identifies
``diagonal" interactions of the vibration with local charge densities on the donor and acceptor sites,
the ``off-diagonal" term $\lambda_{o}$ allows for vibration-induced hopping between local sites.

To understand signatures of transport mechanisms in the thermopower
we now consider three different cases:

(i) Model 1; coherent transport.
Considering Eq. (\ref{eq:Hda}), we eliminate the coupling of electrons to the vibrational model
$\lambda_{o}=0$, $\lambda_{di}=0$. We then reach a noninteracting exactly solvable Hamiltonian
which administers coherent-elastic electron dynamics.

(ii) Model 2; complete decoherence. We allow for pure dephasing but deny energy exchange processes
between electrons and the vibration.
%
In our simulations we only consider complete decoherence, achieved
by erasing interference terms from the transmission function as we explain below Eq. (\ref{eq:Tdeph}).


(iii) Model 3, inelastic transport. In this case electrons cross the molecule assisted by the vibration.
We reach this limit by setting $\Delta=0$ and $\lambda_{di}=0$ in Eq. (\ref{eq:Hda}).
This model can describe transport in e.g., the DMBP molecule as DFT calculations \cite{Markussen} provide $\Delta =0.03$ eV,
vibrational energy $\hbar\omega_0=5$ meV, corresponding to the torsional mode around the central bond
with $\lambda_{di}=0$, $\lambda_{o}=1$ and
$\alpha=9.4\hbar\omega_0$, a large nonperturbative value.
Analysis of the current in this system in the temperature range $T=50-300$ K
revealed that it was overwhelmingly dominated by the inelastic component \cite{Markussen}.
%
Note the following relation between the AO and MO representations,
\bea
&&\alpha(c_d^{\dagger}c_a+c_a^{\dagger}c_d) (b_0^{\dagger}+b_0)
\nonumber\\
&& \rightarrow
\alpha(c_2^{\dagger}c_2 - c_1^{\dagger}c_1)  (b_0^{\dagger}+b_0).
\label{eq:int}
\eea
Thus, a vibration-assisted electron hopping term in the AO picture (model 3) translates to the interaction
of the vibration with the difference of
charge densities on the molecular orbitals.
While we could have
adopted the off-diagonal model (\ref{eq:int}) as our starting point, we preferred to provide the more general interaction form,
Eqs. (\ref{eq:HM}) and (\ref{eq:Hda}),
to clarify the relation between the AO and MO representations in more general cases.


\begin{figure}[htbp]
{\hbox{\epsfxsize=80mm \hspace{0mm}\epsffile{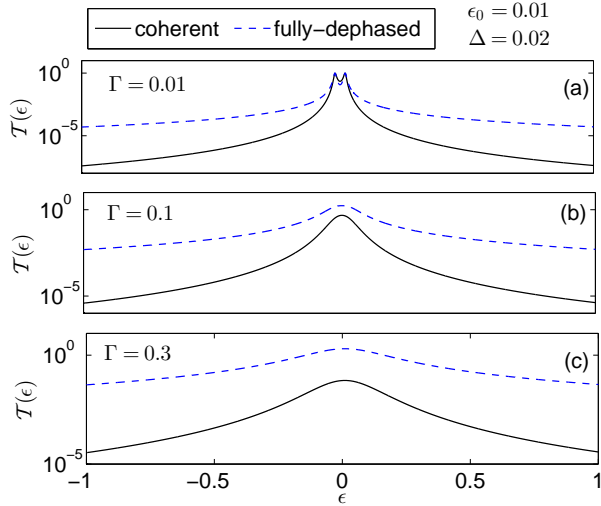}}}
\caption{\normalsize{
Transmission function (\ref{eq:Tdes}) in the coherent model (full)
and the overly dephased case of Eq. (\ref{eq:Tdeph}) (dashed)
using $\Delta=0.02$ eV and $\epsilon_0=0.01$ eV
with the hybridization energy $\Gamma$=0.01, 0.1, 0.3 eV, as indicated in the panels.
}}
\label{trans}
\end{figure}

\section{Computational approaches}
\label{Smethod}

\subsection{Coherent and dephased transport: Landauer-B\"uttiker formalism}
\label{methodL}

{\it Model 1.}
We cross-out electron-vibration interaction terms in Eq. (\ref{eq:Hda}). Electrons transfer the system
elastically and coherently, and the problem can be handled within the Landauer formalism \cite{land},
arriving at the steady-state charge current
\bea
I_e=\frac{e}{2\pi\hbar}\int_{-\infty}^{\infty} 
 \mathcal{T}(\epsilon)\big[f_L(\epsilon) - f_R(\epsilon)\big]d\epsilon.
\label{eq:curr}
\eea
The Fermi-Dirac function $f_{\nu}(\epsilon) = \big[e^{(\epsilon - \mu_{\nu})/k_BT_{\nu}} + 1\big]^{-1}$
provides the charge distribution in the metallic reservoirs $\nu=L,R$, maintained
at the chemical potential $\mu_{\nu}$ and temperature $T_{\nu}$.
The transmission function $\mathcal T(\epsilon)$ can be obtained from the Green's function formalism.
In the linear response regime we Taylor-expand the expression for the charge current around the equilibrium Fermi energy
$\epsilon_F$ and the temperature $T$.
%
%
We then identify the electronic conductance $G=I/\Delta V$, $\Delta \mu=e\Delta V$,
 and the thermopower, the Seebeck coefficient $S\equiv -\frac{\Delta V}{\Delta T}|_{I_e=0}$, by
\bea
G&=&
\frac{e^2}{h}
 \int_{-\infty}^{\infty} \mathcal{T}(\epsilon)\left(-\d{f}{\epsilon}\right)
d\epsilon
\nonumber\\
S &=& -\frac{1}{|e|T}
\frac{ \int_{-\infty}^{\infty} \mathcal{T}(\epsilon)(\epsilon - \epsilon_F)(-\d{f}{\epsilon})
d\epsilon}
{\int_{-\infty}^{\infty} \mathcal{T}(\epsilon)(-\d{f}{\epsilon})d\epsilon}.
\eea
%
The transmission coefficient $\mathcal{T}(\epsilon)$ is obtained
from the Green's function $\mathcal{G}$ and the hybridization matrices $\hat{\Gamma}^{L/R}$
using a standard procedure,
%
$\mathcal{T}(\epsilon) = \Tr[\hat{\Gamma}^{L}(\epsilon)\mathcal{G}(\epsilon)\hat{\Gamma}^{R}(\epsilon)\mathcal{G}^{\dagger}(\epsilon)]$.
Expressions for model 1 can be looked up in the literature, see e.g., Ref. \cite{thoss1}, with
\bea
&&\mathcal{T}(\epsilon)= \frac{(\Gamma/2)^2}{|(\epsilon - \epsilon_0 - \Delta) + i\Gamma/2 |^2}
\nonumber\\
 &&+ \frac{(\Gamma/2)^2}{|(\epsilon - \epsilon_0 + \Delta) + i\Gamma/2 |^2}
\nonumber \\
&&- \frac{1}{2}\Re\bigg[\frac{\Gamma^2}
{(\epsilon - \epsilon_0 - \Delta + i\Gamma/2)(\epsilon - \epsilon_0 + \Delta - i\Gamma/2)}\bigg].
\label{eq:Tdes}
\eea
Here $\Re$ denotes real part. The hybridization energy is defined as
\bea
\Gamma_{\nu}(\epsilon) = 2\pi\sum_{k\in\nu}\abs{\upsilon_k}^2\delta(\epsilon - \epsilon_k).
\label{eq:Ga}
\eea
In our calculations below it is taken as an energy-independent parameter,
identical at the two contacts, $\Gamma=\Gamma_{\nu}$.
The transmission function (\ref{eq:Tdes}) can be also organized as follows
\bea
\mathcal{T}(\epsilon)&=&
\frac{\Gamma^2\Delta^2} { [(\epsilon-\epsilon_0+\Delta)^2+(\Gamma/2)^2]    [(\epsilon-\epsilon_0-\Delta)^2+(\Gamma/2)^2] }.
\nonumber\\
\label{eq:Tdes2}
\eea
This expression
immediately reveals that the current diminishes when $\Delta=0$, a complete destructive interference
in the language of MO.


{\it Model 2.} We now consider elastic dephasing, yet
do not permit inelastic scatterings. 
We reach this limit by dropping interference terms from the transmission function (\ref{eq:Tdes}),
a scheme proposed in Ref. \cite{thoss1}. 
We then arrive at the additive Lorentzian  form
%
\bea
\mathcal{T}_{deph}(\epsilon)=  \sum_{m=\pm}
\frac{(\Gamma/2)^2}{(\epsilon - \epsilon_0 +m \Delta)^2
+ \Gamma^2/4},
\label{eq:Tdeph}
\eea
where the sum over two independent functions indicates on non-interfering transmission pathways for electrons.
This expression can be justified by employing the technique of B\"uttiker's dephasing probe \cite{Buttiker-probe}.
In this approach, a probe reservoir is attached to each molecular orbital and we
apply the dephasing probe condition, demanding that the net current from the probe to the molecule is zero
within each energy component. We then
arrive at the transmission function  (\ref{eq:Tdeph})
in the limit of strong molecule-probe coupling, only with additional probe-induced broadening terms.

Fig. \ref{trans} displays the transmission functions (\ref{eq:Tdes}) and (\ref{eq:Tdeph}),
manifesting the dramatic effect of destructive QI upon increasing the hybridization energy $\Gamma$
to the leads. Recall that when $\Gamma>\Delta$ the molecular orbitals become quasi-degenerate. In
this situation model 2 provides $\mathcal T_{deph}(\epsilon_F)\xrightarrow{\Gamma\gg \Delta} 2$,
whereas in model 1,  $\mathcal T(\epsilon_F)\xrightarrow{\Gamma\gg \Delta} 0$.
These results immediately translate to the electronic conductance at low temperatures.

\subsection{Inelastic transport: Master equation}
\label{methodK}

{\it Model 3.}
As explained in Sec. \ref{Smodel}, we
simplify considerably Eq. (\ref{eq:Hda})
by taking  $\Delta=0$ and $\lambda_{di}=0$,
leaving out only vibration-assisted electron transmission processes.
We study this case by
employing a master equation treatment
as detailed in our recent work \cite{SF}, referred next to as a ``kinetic equation" method.
This approach is perturbative in the electron-vibration interaction $\alpha$ but exact to all orders in the electronic
hybridization energy $\Gamma$. The Born-Markov approximation, lying at the heart of this approach,
assumes that relaxation processes in the leads
are fast enough to erase electronic coherences in the metals.
The secular approximation
ensures that no vibrational coherences persist.
Although the approximations involved here seem quite restrictive,
a comparison to numerically exact path-integral simulations confirmed that
at weak electron-vibration coupling ($\alpha/\hbar\omega_0\lesssim1$), the kinetic equation correctly reproduced
transport characteristics in this donor-acceptor junction \cite{INFPIy}.
Quite interestingly, even at stronger coupling $\alpha/\hbar\omega_0=3$
the kinetic equation provided correct-qualitative results for the junction's current-voltage characteristics \cite{INFPIy}.

The kinetic equation method has been discussed in great details in Ref. \cite{SF}
and here we only include its resulting expressions for the charge current.
In the case of a harmonic (H) vibrational mode, it is given by 
\bea
I_e^{H}= 2e\frac{ k_u^{L\rightarrow R}k_d^{L\rightarrow R} -
k_u^{R\rightarrow L}k_d^{R\rightarrow L}}
{ k_d^{L\rightarrow R} + k_d^{R\rightarrow L}- k_u^{L\rightarrow R} -k_u^{R\rightarrow L}}.
\label{eq:IeH}
 \eea
%
The rate constants describe transitions between vibrational levels in the harmonic oscillator,
driven by the electronic environments $\nu,\nu'=L,R$,
%
%
%
\begin{widetext}
\bea
k_{d}^{\nu\rightarrow \nu'} &=& \frac{1}{2\pi\hbar}\int_{-\infty}^{\infty} d\epsilon
f_{\nu}(\epsilon)(1 - f_{\nu'}(\epsilon+\hbar\omega_0)){J}_{\nu}(\epsilon){J}_{\nu'}(\epsilon+\hbar\omega_0)
\nonumber\\
k_{u}^{\nu\rightarrow \nu'} &=& \frac{1}{2\pi\hbar}\int_{-\infty}^{\infty} d\epsilon
f_{\nu}(\epsilon)(1 - f_{\nu'}(\epsilon-\hbar\omega_0)){J}_{\nu}(\epsilon){J}_{\nu'}(\epsilon-\hbar\omega_0)
\nonumber \\
\label{eq:rates}
\eea
\end{widetext}

These rate constants are given in terms of spectral density functions,
\bea
J_L(\epsilon)&=&\frac{\alpha}{2}\frac{\Gamma_L(\epsilon)}{(\epsilon-\epsilon_0)^2+(\Gamma_L(\epsilon)/2)^2}
\nonumber\\ J_R(\epsilon)&=&
\frac{\alpha}{2}\frac{\Gamma_R(\epsilon)}{(\epsilon-\epsilon_0)^2+(\Gamma_R(\epsilon)/2)^2},
\label{eq:spec}
\eea
with the hybridization energy $\Gamma_{\nu}(\epsilon)$ defined in Eq. (\ref{eq:Ga}).
As mentioned above, in simulations we take $\Gamma=\Gamma_{\nu}(\epsilon)$, an energy independent constant.

The spectral functions (\ref{eq:spec}) represent electronic density of states at the two metals (multiplied by $\alpha/2$),
obtained once absorbing the $d$ ($a$) levels in the $L$ ($R$) lead.
The rates in Eq. (\ref{eq:rates}) are nonzero as long as the overlap between these functions, shifted by $\hbar\omega_0$ due
to energy absorbed (emitted) by electrons from (to) the harmonic oscillator,
is non-negligible.
The Fermi-Dirac functions within the integrands ensure that electrons hop from occupied to empty states.

Equation (\ref{eq:IeH}) provides the current under the assumption that the special vibrational mode is isolated
from other vibrational degrees of freedom. The formalism developed in Ref. \cite{SF}
can be extended to include dissipation mechanisms of this special mode. This is achieved
by attaching it (bilinear coupling) to a secondary bath of harmonic oscillators which do not directly couple to electron transfer
processes in the junction  \cite{SF,INFPIy}.
These secondary harmonic modes are assumed to construct a thermal bath maintained at thermal equilibrium
at temperature $T_{ph}$. The electronic and phononic temperatures may differ, but in this work we assume
them to be the same.
It can be shown that at weak (harmonic mode-secondary bath) coupling
phonon bath-induced relaxation and excitation rate constants satisfy
\bea
k_d^{ph}=\frac{\Gamma_{ph}}{\hbar}[n_{ph}(\omega_0)+1], \,\,\,
k_u^{ph}=\frac{\Gamma_{ph}}{\hbar}n_{ph}(\omega_0),
\eea
with the Bose-Einstein distribution function $n_{ph}(\omega_0)=[e^{\hbar\omega_0/k_BT_{ph}}-1]^{-1}$
and the coupling energy $\Gamma_{ph}$, evaluated at the mode frequency $\omega_0$.
The charge current in this dissipative harmonic mode (H-D) model
follows a compact form,
\bea
I_e^{H-D}= e\frac{
(k_u^{L\rightarrow R } - k_u^{R\rightarrow L })k_d   + (k_d^{L\rightarrow R } - k_d^{R\rightarrow L })k_u}
{k_d-k_u},
\nonumber\\
\label{eq:IeHD}
\eea
with $k_{d,u}$ comprising  both phonon and electron induced rate constants,
$k_d=k_d^{ph}+ k_{d}^{L\rightarrow R}+ k_{d}^{R\rightarrow L}$,
$k_u=k_u^{ph}+ k_{u}^{L\rightarrow R}+ k_{u}^{R\rightarrow L}$.

The kinetic equation as developed in Ref. \cite{SF} can further direct an anharmonic vibrational mode.
We examine vibrational anharmonicity by truncating the harmonic manifold to retain only two states.
In this anharmonic (A) limit, charge current between the metals obeys \cite{SF}
\bea
I_e^{A} =
2e\frac{k_u^{L\rightarrow R}k_d^{L\rightarrow R} -
k_u^{R\rightarrow L}k_d^{R\rightarrow L} }
{k_d^{L\rightarrow R} + k_d^{R\rightarrow L} + k_d^{R\rightarrow L} + k_u^{R\rightarrow L}}.
\label{eq:IeA}
\eea
Closed-form expressions for the conductance and the thermopower
can be obtained by Taylor-expanding the charge current
in $\Delta V$ and $\Delta T$, to extract linear response coefficients.
However, a direct, brute-force, numerical procedure has been proved easier to implement and converge:
We evaluate the current at a series of small $\Delta T$, then search for the voltage bias $\Delta V$ which nullifies
the current. We plot these voltages as a function of $\Delta T$, verify a linear relation, and obtain
the thermopower from the slope.

Several remarks are now in order:
(i) The kinetic approach is perturbative in $\alpha$, the electron-vibration coupling energy,
and it provides the scaling  $I_e\propto \alpha^2$.
Values obtained for the electronic conductance can thus be immediately scaled to
account for stronger or weaker electron-vibration interaction energies. Furthermore,
the thermopower in the kinetic treatment does not depend on $\alpha$, as this factor cancels down.
(ii)
It is interesting to point out on the similarity between the transmission function (\ref{eq:Tdes2}) in the coherent model
and the integrands in expressions for the rate constants, Eq.  (\ref{eq:rates}) of the kinetic scheme.
The transmission function (\ref{eq:Tdes2}) is a product of shifted Lorentzian functions reflecting
quantum interference effects between electrons in different MOs. We could also understand this behavior by defining
weighted density of states around molecular orbitals,
$K(\epsilon\pm\Delta)\equiv\frac{\Gamma} {(\epsilon-\epsilon_0\pm\Delta)^2+(\Gamma/2)^2}$,
providing the transmission  $\mathcal T(\epsilon)=\Delta^2K(\epsilon+\Delta)K(\epsilon-\Delta)$.
The rate constants in the kinetic approach have a similar structure: We multiply
electronic densities of states (\ref{eq:spec}) shifted by $\omega_0$ 
to construct electron transfer rates between the leads. The additional factor $\alpha^2$ connects
the two spectral functions, as electron transfer is assisted by an interaction with a vibrational mode.

Given this connection between models 1 and 3,
transport characteristics of the coherent model can be meaningfully contrasted to the kinetic equation
by using density of states $K(\epsilon)$ (model 1)
comparable to the spectral function $J(\epsilon)$ (model 3).

\begin{figure}[htbp]
\vspace{0mm} {\hbox{\epsfxsize=85mm \hspace{0mm}\epsffile{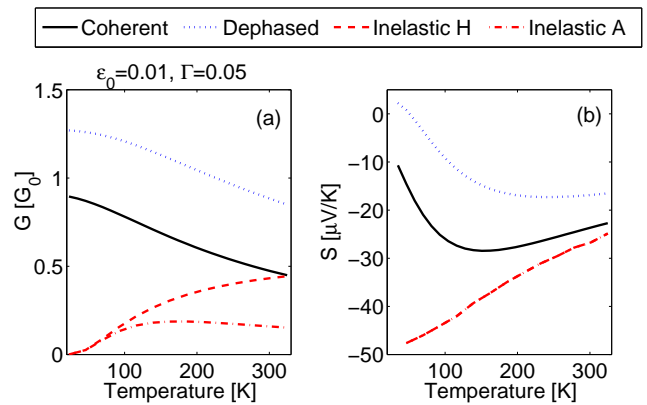}}}
\caption{\normalsize
{
Resonant transport with weak interference effects.
(a) Electrical conductance and (b) thermopower in the
coherent case, model 1 (solid); fully dephased case, model 2 (dotted);
inelastic model 3 assuming  harmonic  (H, dashed)
or an anharmonic two-state mode (A, dashed-dotted).
The inelastic lines overlap in panel (b).
Parameters (in eV) are $\epsilon_0=0.01$, $\Gamma=0.05$ and $\Delta =0.02$ in model 1 and 2,
$\hbar\omega_0=0.02$ and $\alpha=0.02$ in model 3.
}}
\label{Fig1}
\end{figure}

\begin{figure}[htbp]
\vspace{0mm} {\hbox{\epsfxsize=85mm \hspace{0mm}\epsffile{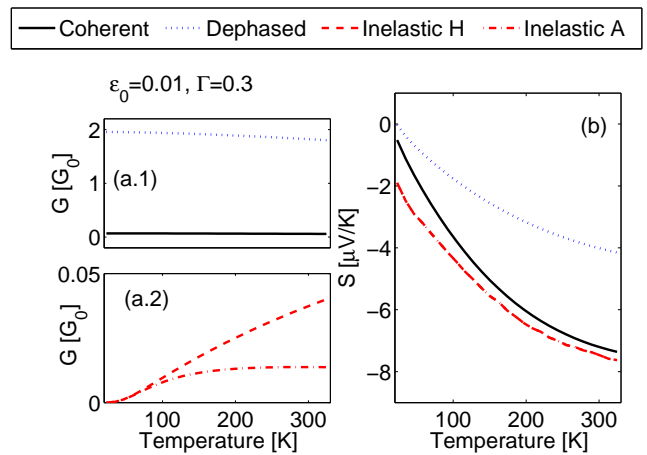}}}
\caption{\normalsize
{Resonant transport with strong interference effects.
(a.1)-(a.2) Electrical conductance and (b) thermopower in
model 1 (solid); model 2 (dotted);
model 3 with a harmonic  (H, dashed)
or an anharmonic two-state mode (A, dashed-dotted).
These last two lines overlap in panel (b).
Parameters (in eV) are  the same as in Fig. \ref{Fig1}, besides $\Gamma=0.3$.
}}
\label{Fig2}
\end{figure}

\begin{figure}[htbp]
\vspace{0mm} {\hbox{\epsfxsize=85mm \hspace{0mm}\epsffile{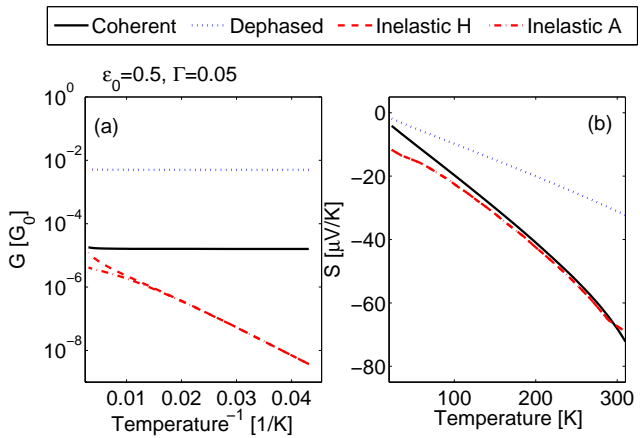}}}
\caption{\normalsize
{Off-resonant transport.
(a) Electrical conductance as a function of the inverse temperature 
and (b) thermopower as a function of temperature for
model 1 (solid); model 2  (dotted);
model 3 with a harmonic (H, dashed)
and an anharmonic two-state mode (A, dashed-dotted).
These last two lines overlap in panel (b).
Parameters (in eV) are $\epsilon_0=0.5$,  $\Gamma=0.05$, and $\Delta =0.02$ in model 1 and 2,
and $\hbar\omega_0=0.02$ and $\alpha=0.02$ for model 3.
}}
\label{Fig3}
\end{figure}


\section{Results}
\label{Sresult}

We study the electrical conductance and the Seebeck coefficient of model (\ref{eq:Hda}) for coherent, dephased and
vibration-assisted electrons, as described in Sec. \ref{Smodel}, by using the techniques explained in Sec. \ref{Smethod}.

We consider two sets of parameters.
In section \ref{R1} we look for fingerprints of transport mechanisms in $G$ and $S$ using
model parameters.
In section \ref{R2} we study the 2,2'- dimethylbiphenyl (DMBP) molecule, the particular vibration
corresponds to a torsion mode of the two benzene rings.
DFT calculations
indicate that the electron-vibration interaction in this molecule is non-perturbative,
$(\alpha/2)/\hbar\omega_0\sim 10$. \cite{Markussen}
However, in a recent study we confirmed that the kinetic treatment, as described in Section \ref{methodK},
provides accurate qualitative features for
the current, even beyond the strict weak coupling limit \cite{INFPIy}.
Further, in Ref. \cite{Markussen} it was noted that the self-consistent Born Approximation
feasibly converged (after a single iteration), to give the vibration-assisted
charge current in the DMBP junction. Given this facile convergence,
one may argue that the perturbative treatment is meaningful
even beyond its strict range.
Furthermore, while the magnitude of the charge current is overestimated in
the kinetic approach (the scaling $\alpha^2$ should be saturated at strong coupling) \cite{INFPIy},
the expression for the thermopower is independent of $\alpha$.
Given these arguments, we believe that the kinetic treatment, known to provide correct
qualitative features for the current, yields correct qualitative and quantitative answers for $S$,
even in nonperturbative electron-vibration cases.

\subsection{Expressions for $G$ and $S$}
\label{R1}

We consider quasi-degenerate situations and contrast resonance and off-resonance settings.
In resonant situations we use $\epsilon_0 = 0.01$ eV
and $\Delta = 0.02$ eV, the Fermi energy  $\epsilon_F$ is set as zero. In this configuration,
the two orbitals are placed at $\epsilon_0 - \Delta = -0.01$ eV and $\epsilon_0 + \Delta = 0.03$ eV;
In off-resonant situations we settle on $\epsilon_0 = 0.5$ eV
and $\Delta = 0.02$ eV. In this case the two molecular orbitals are situated above the Fermi energy, 
  $|\epsilon_0\pm\Delta|>\Gamma$.
%
In the inelastic model 3 we use $\hbar\omega_0=0.02$ eV for the vibrational frequency.
 Our observation below are not specific to the special case $\Delta=\hbar\omega_0$,
in fact, at large hybridization our results for $S$ are insensitive to the exact value assumed for these parameters, as we discuss below.
The electron-vibration interaction is set at
$(\alpha/2)=0.01$ eV, but this parameter only scales the conductance
in the kinetic approach $G\propto\alpha^2$ and it does not
affect the behavior of the thermopower under our treatment.
In this section we consider either a harmonic mode (inelastic H), or an anharmonic two-state
case (inelastic A), obtained from Eqs. (\ref{eq:IeH}) and (\ref{eq:IeA}) respectively.
We vary the temperature of the electronic baths
in the range $T=50-300$ K. The hybridization energies are modified in a broad range, $\Gamma=0.01-0.5$ eV.
When $\Gamma/\Delta\ll 1$, the two orbitals independently conduct and quantum interference effects are negligible.
In the opposite $\Gamma/\Delta>1$ limit the levels become quasi-degenerate and destructive interference
features are apparent in model 1.
We plot $G$ in units of the conductance quantum per mode per spin specie, $G_0=e^2/h$.
The thermopower $S$ is plotted in units of $\mu V/K$.

{\it Temperature dependence.}
In Fig.  \ref{Fig1}  we display the  conductance and the thermopower vs. the electronic temperature.
Since $\Gamma$ is not large, the coherent and dephased models similarly behave.
Under inelastic effects (model 3) $G$ increases with temperature,
contrasting the behavior of models 1 and 2.
The reason for this enhancement is the following: In model 3 electrons transfer the system only by
exchanging energy $\hbar\omega_0$ with
the vibrational mode. Upon heating the electron bath we increase electron occupation of levels with
$|\epsilon_k|>\hbar\omega_0$, opening additional inelastic channels.
While the thermopower is sensitive to the underlying transport mechanism at low temperatures $T<200$ K,
at higher temperature, $T=200-300$ K we {\it cannot} well differentiate between the different cases
 merely by inspecting the thermopower, see Fig. \ref{Fig1}(b).

In Fig. \ref{Fig2} we increase the hybridization energy and the orbitals can now be considered as quasi-degenerate.
As a result, the conductance in model 1 is significantly suppressed due to destructive interference,
compared to the fully-dephased limit which approaches the maximal value $2 G_0$.
Both $G_1$ and $G_2$ decrease with temperature (the subscript identifies the model). In contrast,
model 3 behaves as $G_3\propto e^{-\beta E_a}$, with $E_a=18$ meV, close to the vibrational frequency,
see the discussion following Fig. \ref{Fig3}.

Given these differences in electronic conduction,
it is intriguing to note that the three models provide close results
for the thermopower. Particularly, $S_1\sim S_3$.
Furthermore, here and in other cases we note that the thermopower does not depend on the mode harmonicity/anharmonicity.
We thus conclude that in the present model the thermopower serves as an excellent tool for identifying
the energetics of the junction, but it cannot readily uncover mechanisms of electron dynamics in the system.

We continue to off-resonance situations with $\epsilon_0=0.5$, see Fig. \ref{Fig3}.
The electronic conductance again clearly manifests significant sensitivity to transport mechanisms:
In the coherent case, as well as under pure dephasing,  $G\propto T^{0}$. In contrast,
under inelastic effects $\log G\propto \hbar\omega_0/T$, as expected for an activated process \cite{active}.
The thermopower is again indifferent to transport mechanisms, showing here a clear linear trend, $S\propto  T$,
in all three models, with either an harmonic or anharmonic local mode.

We now derive a closed expression for $S$ in the off-resonant regime $|\epsilon_0|>\Gamma$. 
Assuming that the transmission function is smooth near the Fermi energy (set as $\epsilon_F=0$),
the current can be analyzed with a Sommerfeld expansion.
At low enough temperatures, $k_BT<|\epsilon_0|$, one obtains the (standard) result \cite{datta}
\bea
S=-
\frac{\pi^2k_B^2T}{3|e|} \frac{\partial \ln \mathcal T(\epsilon)}{\partial \epsilon} \Big| _{\epsilon=\epsilon_F}.
\label{eq:SS}
\eea
We work out this expression with
 the transmission function (\ref{eq:Tdes2}) or (\ref{eq:Tdeph}).
When the levels are quasi-degenerate, $\Gamma>\Delta$, we get
\bea
S_{quasi}=\frac{\pi^2k_B^2T}{3|e|}  \left[\frac{p}{\epsilon_F-\epsilon_0}\right].
\label{eq:Squasi}
\eea
Here $p=4$ is received for the coherent case. Under full dephasing, when Eq. (\ref{eq:Tdeph}) is employed, $p=2$,
exactly half the coherent value. Equation (\ref{eq:Squasi}) was derived in Ref. \cite{fullerene},
by approximating the transmission function of a fullerene-based junction
to be a sum of independent Lorentzians, a dephased situation in our language.
Our results extend the discussion in Ref. \cite{fullerene}, to allow interference effects between orbitals.

Numerical simulations in Fig. \ref{Fig3} support Eq. (\ref{eq:Squasi}): The thermopower is linear with temperature,
and the slope of the dephased case is a factor of two smaller than the coherent case.
Note that the thermopower is negative when $\epsilon_0-\epsilon_F>0$.
In other words,
when the conductance is dominated by LUMO, LUMO+1 orbitals, $S<0$. In contrast,
when transport is dominated by the HOMO, HOMO-1 orbitals, $S>0$.

Eq. (\ref{eq:Squasi}) is one of our main results, valid for  $|\epsilon_0|>\Gamma>T,\Delta$.
When the HOMO and LUMO orbitals are
separated by a large energy gap, $2\Delta > |\epsilon_F-\epsilon_0|>T,\Gamma$,
a different expression holds. 
In this  ``gapped"  situation, resonance features are separated in energy and interference effects are non-significant.
Using Eq. (\ref{eq:Tdes}), we reach a simple expression for the thermopower  \cite{datta},
\bea
S_{gap}=-\frac{2\pi^2k_B^2T}{|e|} \left[ \frac{\epsilon_F-\epsilon_0}{\Delta^2}\right].  
\label{eq:SLarge}
\eea
It indicates that when the HOMO level is aligned closer to the Fermi energy than the LUMO level,
i.e., when
$\epsilon_0>\epsilon_F$, $S$ attains a positive sign, and vice versa.

\begin{figure}[htbp]
\vspace{0mm} {\hbox{\epsfxsize=85mm \hspace{0mm}\epsffile{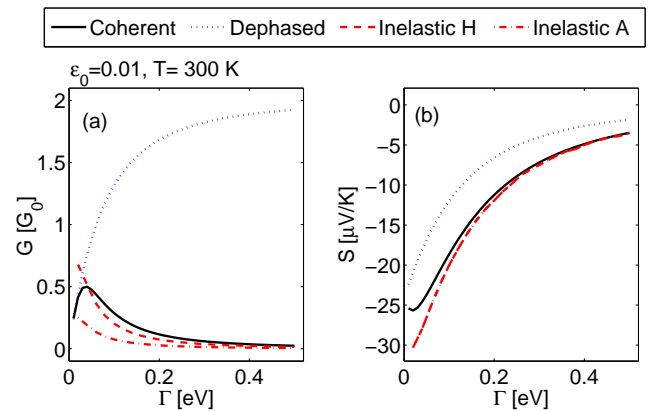}}}
\caption{\normalsize
{
(a) Electronic conductance
and (b) thermopower as a function of the hybridization energy $\Gamma$ in
model 1 (solid); model 2 (dotted);
model 3 with a harmonic mode (H, dashed)
and an anharmonic two-state mode (A, dashed-dotted).
These last two lines overlap in panel (b).
Parameters (in eV) are $\epsilon_0=0.01$,  and $\Delta =0.02$ in model 1 and 2,
and $\hbar\omega_0=0.02$ and $\alpha=0.02$ for model 3, $T=300$ K in all simulations.}}
\label{Fig4}
\end{figure}

\begin{figure}[htbp]
\vspace{0mm} {\hbox{\epsfxsize=85mm \hspace{0mm}\epsffile{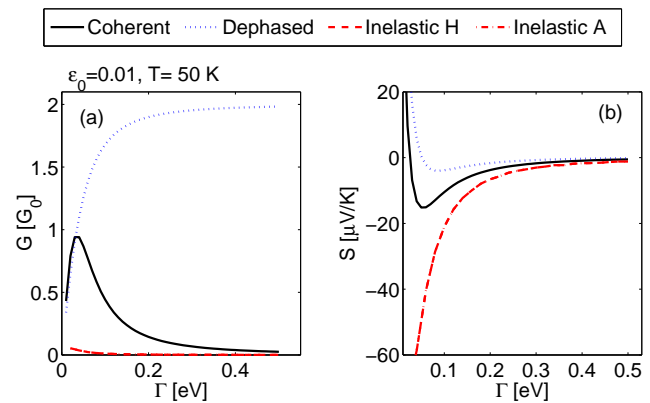}}}
\caption{\normalsize
{
Same as Fig. \ref{Fig4}, with $T=50$ K.
}}
\label{Fig5}
\end{figure}

\begin{figure}[htbp]
\vspace{0mm} {\hbox{\epsfxsize=85mm \hspace{0mm}\epsffile{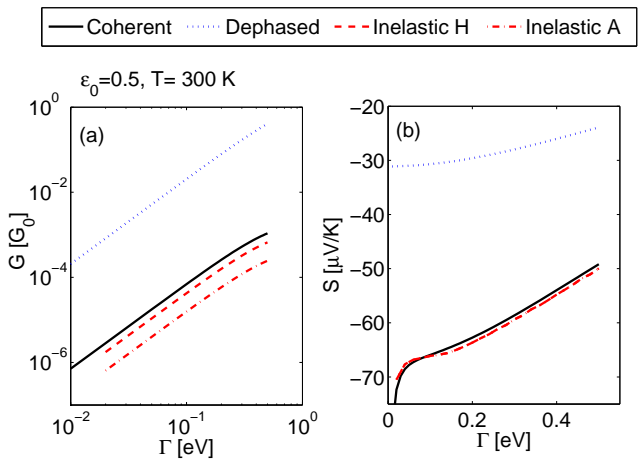}}}
\caption{\normalsize
{Off-resonance situation:
(a) Electronic conductance
and (b) thermopower as a function of the hybridization energy $\Gamma$ in
model 1, coherent transport (solid); model 2, fully dephased electrons (dotted);
model 3 with only inelastic effects using a harmonic mode (H, dashed)
and an anharmonic two-state system mode (A, dashed-dotted).
Parameters (in eV) are $\epsilon_0=0.5$,  and $\Delta =0.02$ in model 1 and 2,
and $\hbar\omega=0.02$ and $\alpha=0.01$ for model 3, $T=300$ K in all simulations.
}}
\label{Fig6}
\end{figure}

{\it Hybridization energy.}
In Fig. \ref{Fig4} we display the conductance and the thermopower as a function of the molecule-metal
hybridization energy.
Comparing model 1 to model 2, we again observe that interference effects become detrimental for the transport
once $\Gamma>\Delta$, when the orbitals become quasi-degenerate.
Beyond that point, the conductance in model 1 decays with $\Gamma$ as $G\propto1/ \Gamma^{2}$.
The inelastic model follows a similar decay function
since increasing $\Gamma$ broadens the electronic density of states (with the donor and acceptor levels absorbed in the leads),
while lowering overlap integrals at relevant resonance energies, see Eq. (\ref{eq:spec}).
In Fig. \ref{Fig4}(b) we show that the coherent and inelastic cases support similar values for the Seebeck coefficient at room temperature.
Substantial differences between the three cases are displayed in Fig. \ref{Fig5}(b), at low enough temperature.
Specifically, the inelastic component monotonically decays with increasing $\Gamma$, while the coherent and dephased cases
show a turnover behavior, with the thermopower switching its sign around $\Gamma=0.05$ eV.
Furthermore, at low temperature the magnitude of the thermopower is significantly enhanced
when $\Gamma$ is small, under the three different mechanisms.
In Fig. \ref{Fig6} we study the behavior of $G$ and $S$ in off-resonance situations. We find
that $G\propto \Gamma^2$ in the three models, and that the thermopower develops in a similar fashion for the three cases.
Simulations of off-resonance situation at lower temperature, $T=50$ K follow similar trends (not shown).


We can now organize some simple relations for the conductance and the thermopower. For a system with quasi-degenerate levels in resonance situations, $\Gamma>\Delta,|\epsilon_0-\epsilon_F|$, we find that
\bea
G/G_0 \propto
\begin{cases}
\frac{\Delta ^2}{\Gamma^2}T^0 & {\rm coherent} \nonumber \\
\left (2-\sum_{\pm}\frac{(\epsilon_F-\epsilon_0\pm\Delta)^2}{(\Gamma/2)^2}\right) T^0 & {\rm dephased} \nonumber\\
\frac{\alpha^2}{\Gamma^2} e^{-\hbar\omega_0/k_BT} & {\rm inelastic} \nonumber \\
\end{cases}
\eea
Our simulations indicate that
the thermopower grows with $T$ in a nonlinear fashion and that it reduces with increasing $\Gamma$. A simple analytical form for $S$ is missing,
but we have consistently confirmed that $S_1\sim S_3=2S_2$ (with the subscript identifying the model).

For a system with quasi-degenerate levels in off-resonance situations, $|\epsilon_0-\epsilon_F\pm\Delta|>\Gamma>\Delta$,
we receive
\bea
G/G_0\propto
\begin{cases}
\frac{\Gamma^2\Delta ^2}{ (\epsilon_F-\epsilon_0)^4}T^0 & {\rm coherent}
\nonumber\\
\frac{\Gamma^2}{(\epsilon_F-\epsilon_0)^2}T^0 & {\rm dephased}
\nonumber\\
\frac{\alpha^2\Gamma^2}{(\epsilon_F-\epsilon_0)^4 
} e^{-\hbar\omega_0/k_BT} & {\rm inelastic}
\end{cases}
\eea
and
\bea
S_{1,3}&\propto&  \frac{k_B^2T}{|e|} \left[\frac{1}{\epsilon_F-\epsilon_0}\right]
\nonumber\\
S_{2}&\sim& S_1/2,
\eea
see Eq. (\ref{eq:Squasi}).
Note that the behavior of $S_3$ was deducted from numerical simulations.

We summarize our observations, regarding the role of different factors on the Seebeck coefficient:

(i) Harmonicity/anharmonicity of the vibrational mode.
In Model 3 electrons transfer the molecule assisted by a specific molecular vibration, harmonic or anharmonic. The electrical
conductance is enhanced when the vibrational mode is harmonic, compared to an
anharmonic mode, since the former supports more conductance channels at high enough temperatures.
In contrast, the Seebeck coefficient is {\it identical} in the two situations, concealing information over
the mode harmonicity/anharmonicity.

(ii) Dissipationless/dissipative mode.
The specific molecular vibration (coupled to electron transfer) may be isolated, or
coupled to a bath of thermalized vibrations.
We examined the role of this dissipation process
on the conductance and the Seebeck coefficient, but did not identify significant effects in this linear response situation.
Allowing the mode to dissipate energy should be imperative to molecular stability
far-from-equilibrium, when the vibrational mode may be (significantly) excited \cite{SF,INFPIy}.


(iii) Enhancement of electrical conductance and Seebeck coefficient.
To maximize both $G$ and $S$, important for enhancing the figure of merit, thus the efficiency of a thermoelectric device,  we suggest a junction with the parameters  $\Gamma\sim \Delta $, see Figs. \ref{Fig4}-\ref{Fig5}.
In this case destructive interference effects are not yet in full play, but $\Gamma$ is large enough 
to allow strong conduction, see Fig. \ref{Fig4}.

(v) Identifying (coherent, dephased, inelastic) transport mechanisms and interference effects in the thermopower.
At low temperatures and small hybridization the different mechanisms may
be separated from thermopower measurements, see Fig. \ref{Fig5}. However,  it is intriguing to note that
when destructive interference dominates coherent dynamics (identified by the comparison between $G_1$ to $G_2$),
the thermopower in the coherent-elastic case behaves similarly to the case with inelastic interactions,
see for example Fig. \ref{Fig3}.

\subsection{The DMBP molecule}
\label{R2}

\begin{figure}[t]
\vspace{-14mm} {\hbox{\epsfxsize=60mm \hspace{15mm}\epsffile{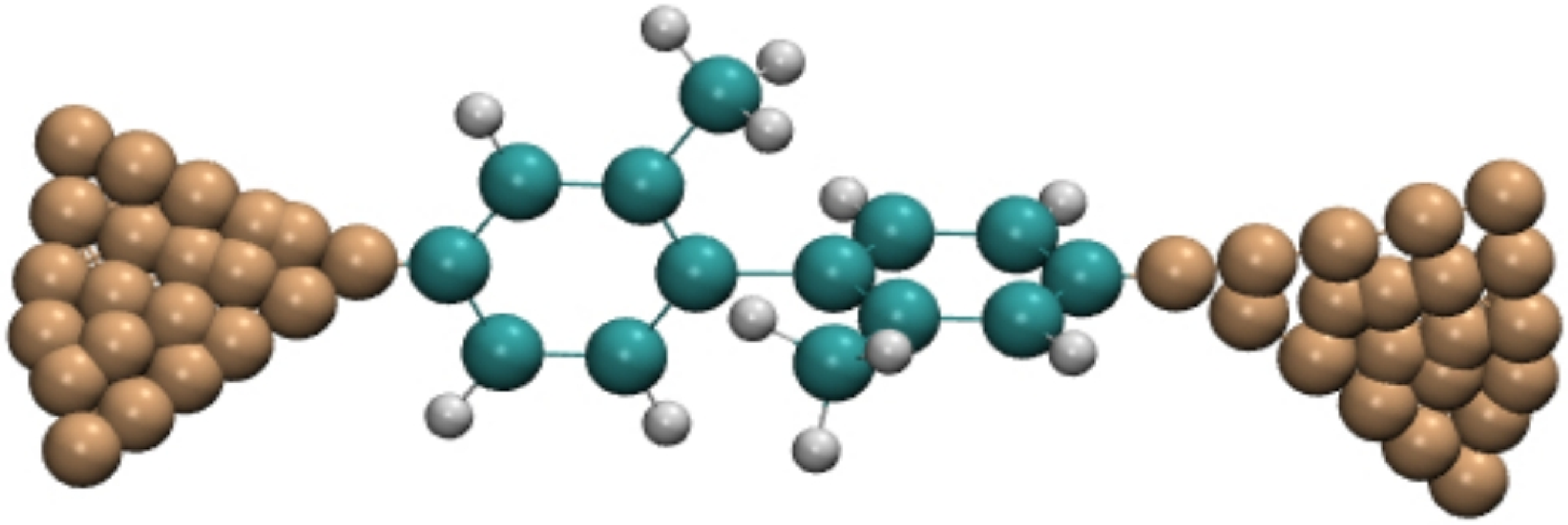}}}
\vspace{-15mm} {\hbox{\epsfxsize=85mm \hspace{0mm}\epsffile{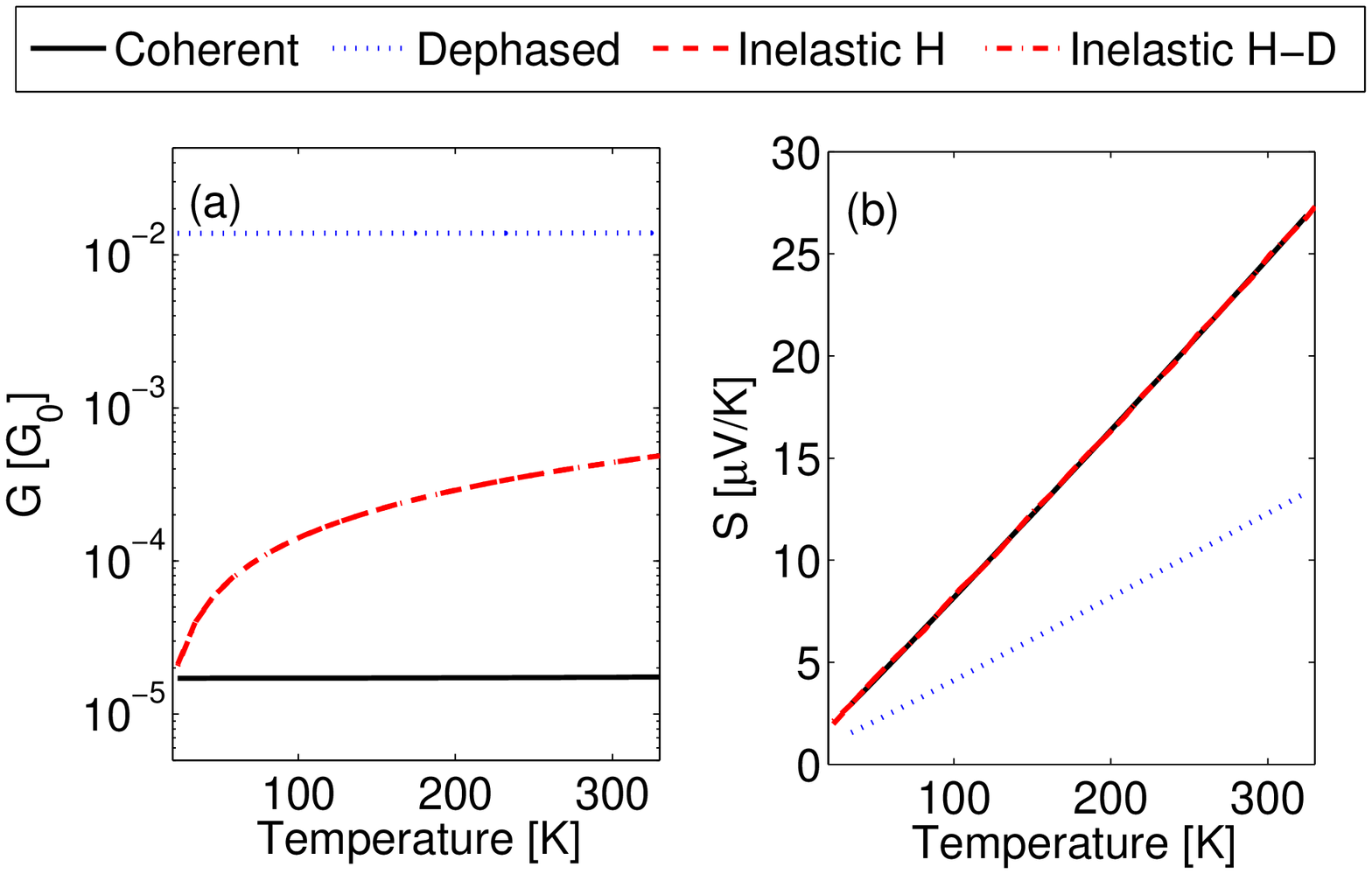}}}
\caption{\normalsize
Top: Pictorial representation of the DMBP molecule between metal leads.
Bottom:
(a) Electrical conductance and (b) thermopower vs. temperature for
the DMBP molecule
for coherent  (full), dephased (dotted),
and inelastic transport. In the latter case the special vibration may
be isolated from additional modes (dashed) or
coupled to a secondary bath (assuming $T_{ph}=T$) with strength $\Gamma_{ph}=1$ meV (dashed-dotted, overlapping with dashed line).
}
\label{FigDMBP1}
\end{figure}

We turn to a specific example and study the conductance and thermopower of the DMBP molecule,
see the top panel in Fig. \ref{FigDMBP1}.
In the AO picture the two benzene rings of the DMBP molecule are oriented almost orthogonal to each other,
with a tilt angle of 81$^o$. As a result, the $\pi$ systems on the benzene rings
are weakly coupled, with a tunneling element $\Delta=0.03$ eV. \cite{Markussen}
DFT calculations provided a torsional vibrational mode of $\hbar\omega_0=0.005$ eV,  strongly coupled
to the electronic degrees of freedom,
$(\alpha/2)/\hbar\omega_0=9.4$.
We also set $\epsilon_0=-1.2$ eV and $\Gamma=0.2$ eV as in Ref. \cite{Markussen}.

Fig. \ref{FigDMBP1} displays the conductance and the thermopower in the DMBP junction as a function of temperature.
We include the full dephasing limit to demonstrate that
interference effects participate in model 1.
The inelastic model is analyzed in two situations: considering only the torsional mode, isolated from
other vibrations (dashed line), or  allowing it to dissipate energy to a secondary harmonic bath,
including other molecular
and environmental modes (dashed-dotted). Our simulations confirm that
since we are considering here low-bias currents, opening an additional
dissipation pathway for the vibrational mode does not affect our results.
We find that the inelastic contribution is thermally activated, and that it dominates electron conduction, see panel (a),
in agreement with Ref. \cite{Markussen}.
In panel (b) we plot the thermopower as a function of temperature;
note that this quantity is non-additive in the three terms (coherent, dephased and inelastic); the inelastic component
with the largest (charge and energy) currents
dictates $S$. It is again interesting to note, as mentioned several times in this work, that
the different mechanisms, coherent and inelastic, support identical behavior for $S$, satisfying Eq. (\ref{eq:Squasi}).

\begin{figure}[htbp]
\vspace{0mm} {\hbox{\epsfxsize=85mm \hspace{0mm}\epsffile{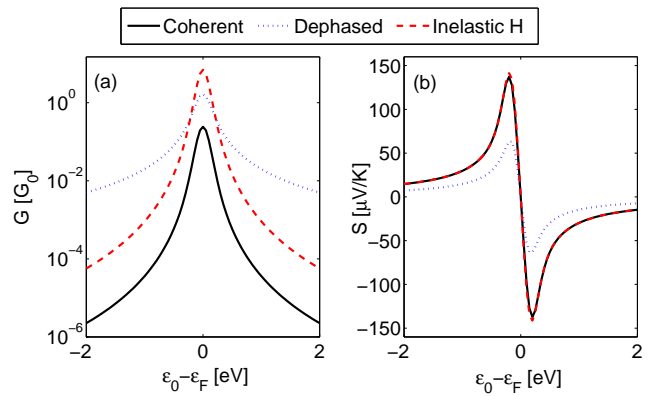}}}
\caption{\normalsize
Gate control over (a) electrical conductance and (b) thermopower
of the DMBP molecule for
coherent (full), dephased (dotted),
and inelastic electrons (dashed),
$T=300 $ K.
}
\label{FigDMBP2}
\end{figure}

Following a recent experimental demonstration of gate control of thermoelectricity in molecular junctions \cite{Reddy14},
in Fig. \ref{FigDMBP2}  we tune the position of the two orbitals $\epsilon_0$ relative to the Fermi energy $\epsilon_F$,
while keeping $\Delta$ fixed.
Given the spatial symmetry of the junction, the conductance and the thermopower are symmetric around
the Fermi energy.
Note that the inelastic contribution reaches several $G_0$ in resonance; the harmonic mode allows for multiple
channels since $k_BT>\hbar\omega_0$.
The thermopower further displays an interesting behavior, with a maximum showing up at $\epsilon_0=\pm 0.2$ eV,
followed by a decay at stronger gating,  $S\propto (\epsilon_0-\epsilon_F)^{-1}$.

\section{Conclusions}
\label{Ssummary}

We studied the behavior of the electrical conductance and the
Seebeck coefficient in a class of molecular junctions with quasi-degenerate molecular orbitals
which potentially suffer from strong destructive interference effects.
We performed our analysis
considering different-limiting transport mechanisms for electrons: coherent, fully dephased,
and inelastic - exchanging energy with a particular (isolated or thermalized) vibration.
The coherent and the fully dephased cases were studied at the level of the Landauer formalism.
Inelastic effects were accounted for by using a
Born-Markov second-order quantum master equation, perturbative in the electron-vibration interaction.
%
%

We are now able to address the question posed in the title of this work: Can the Seebeck coefficient identify
quantum interference in molecular conduction? Our calculations indicate that
room-temperature measurements of the thermopower of molecules with quasi-degenerate conducting orbitals of different (gerade, ungerade)
symmetry (see the geometry of Fig. \ref{Fig2}) cannot provide
decisive information on the role played by quantum interference effects in electron dynamics:
The Seebeck coefficient displays similar characteristics for coherent electrons, and when electrons scatter inelastically by a
molecular vibration, leaving out only structural information.
This observation could explain the significant success of the Landauer formalism in describing
thermopower measurements, even in resonance situations when vibrational modes may play significant roles in electron dynamics \cite{Reddy14}.

Other contributions of this work include the construction of simple expressions for the electrical conductance and the thermopower
under different mechanisms in the limits of resonant or off-resonant transport.
Specifically, we derived a simple analytical form for the thermopower in the case
of quasi-degenerate levels in off-resonant settings, complementing the case with largely
separated HOMO-LUMO levels \cite{datta}.
We further showed that while full dephasing recovers electrical conductance in our model, by eliminating
destructive interference effects,
the thermopower of a dephased system is reduced by exactly a factor of two, compared to the coherent limit.

Our findings demonstrate that many-body effects in electron dynamics may not be revealed in the thermopower.
It is of interest to study molecules of other geometries with strong electron-vibration terms, take into account both coherent and vibration-assisted tunneling terms with experimentally-relevant values, and examine the behavior of the electrical conductance and the thermopower under
different conditions: gating, chemical composition and contact energy.

\begin{acknowledgments}
This work was funded by the Natural Sciences and Engineering
Research Council of Canada and the Canada Research Chair Program.
The work of LS has been supported by the Early
Research Award of DS and the Ontario Graduate Scholarship.
W.J.C acknowledges support from the NSERC USRA program.
We thank A. F. Izmaylov for discussions and DFT data.
\end{acknowledgments}



{}


\begin{thebibliography}{5}
\bibitem{active}
Note that the activation energy in our kinetic model is given $\omega_0$, rather than the electronic energy $\epsilon_0$ which acts as
the activation barrier in other situations \cite{SegalET}.
\end{thebibliography}
\end{document}